\documentclass[     twocolumn,     seceq ]{jpsj3}
\usepackage{graphicx}% Include figure files
\usepackage{dcolumn}% Align table columns on decimal point
\usepackage{bm}% bold math
\def\runauthor{F. J. \textsc{Ohkawa} and T. \textsc{Toyama}}
\def\huscup{Published in J. Phys. Soc. Jpn. {\bf 78} (2009) 124707 (1-11) \newline DOI: 10.1143/JPSJ.78.124707 \newline \copyright2009 The Physical Society of Japan}
\begin{document}
\title{% 
Frustrated Electron Liquids in the Hubbard Model%
\thanks{\huscup}
}
\author{Fusayoshi J. \textsc{Ohkawa}\thanks{E-mail: fohkawa@mail.sci.hokudai.ac.jp} and Takahiro \textsc{Toyama}}
%
%\affiliation{}
\inst{Department of Physics, Graduate School of Science, 
Hokkaido University, Sapporo 060-0810, Japan}
\recdate{July 20, 2009; accepted September 24, 2009; published December 10, 2009}
%\received{\today}
%
\abst{
The ground state of the Hubbard model is studied within the constrained Hilbert space where no order parameter exists. 
The self-energy of electrons is decomposed into the single-site and multisite self-energies. The calculation of the single-site self-energy is mapped to a problem of self-consistently determining and solving the Anderson model.
When an electron reservoir is explicitly considered, it is proved that the single-site self-energy is that of a normal Fermi liquid even if the multisite self-energy is anomalous.
Thus, the ground state is a normal Fermi liquid in the supreme single-site approximation (S$^3$A).
In the strong-coupling regime, the Fermi liquid is stabilized by the Kondo effect in the S$^3$A and is further stabilized by the Fock-type term of the superexchange interaction or the resonating-valence-bond (RVB) mechanism beyond the S$^3$A. 
The stabilized Fermi liquid is frustrated as much as an RVB spin liquid in the Heisenberg model. 
It is a relevant {\it unperturbed} state that can be used to study  a normal or anomalous Fermi liquid and an ordered state in the whole Hilbert space by Kondo lattice theory.
Even if higher-order multisite terms than the Fock-type term are considered, the ground state cannot be a Mott insulator.
%an insulator with a complete gap, nor  
It can be merely a gapless semiconductor even if the multisite self-energy is so anomalous that it is divergent at the chemical potential.
A Mott insulator is only possible as a high temperature phase.
}
\kword{ Mott insulator, metal-insulator transition, Hubbard model, single-site approximation, Kondo effect, Kondo lattice, Fermi liquid, RVB, frustration, third law of thermodynamics}
%
% 
%%%%%%%%%%%%%%%%%%%%%%%%%%%%%%%%%%%%%%%%
%\pacs{71.30.+h, 71.10.-w, 71.27.+a, 74.20.-z}
 %
 % 71.10.-w Theories and models of many electron systems
 % 71.10.Ay Fermi-liquid theory and other phenomelological models
 % 71.27.+a Strongly correlated electron systems; heavy fermions
 % 71.30.+h Metal-insulator transitions 
 %     and other electronic transitions
 % 74.20.-z Theories and models of superconducting state
 % 74.90.n, Other topics in superconductors
 %     (restricted to new topics in section 74) 
 % 75.10.-b General theory and models of magnetic ordering
 % 75.10.Lp Band and itinerant models
 % 75.30.Et Exchange and superexchange interactions
 % 75.30.Kz Magnetic phase boundaries
 % 75.30.Gw Magnetic anisotropy
 % 75.50.-y Studies of specific magnetic materials
%%%%%%%%%%%%%%%%%%%%%%%%%%%%%%%%%%%%%%%%
\maketitle

%%%%%%%%%%%%%%%%%%%%%%%%%%%%%%%%%%%%%%%%%%%%
%%%%%%%%%%%%%%%%%%%%%%%%%%%%%%%%%%%%%%%%%%%%
\section{Introduction}
\label{SecIntroduction}
It is an important but longstanding issue to elucidate the nature of the Mott metal-insulator transition.\cite{mott} Although many studies have been performed, the issue is still contentious. The third law of thermodynamics appears to be broken in a Mott insulator, as discussed below in this section. One of the most contentious issues is whether or not 
the ground state can be really a highly degenerate insulator or
the third law can be really broken 
in a physically relevant model, or the Hubbard model with a finite on-site repulsion $U$, which is one of the simplest effective Hamiltonians to study this issue.

According to Hubbard's theory, \cite{Hubbard1,Hubbard2} when the on-site  $U$ is as large as or larger than the bandwidth $W$, a band splits into two subbands;
the subbands and a gap between them are called the upper Hubbard band (UHB), the lower Hubbard band (LHB), and the Hubbard gap, respectively. 
Hubbard's theory can be pictured by a simple argument.
First, consider the case of $U/W=+\infty$. The number of electrons per unit cell is denoted by $n$. When $n=1$, all the unit cells are singly occupied and there is no empty or double occupancy.
The half-filled ground state is the highly degenerate state of having an electron of arbitrary spin at each unit cell; the entropy is $k_{\rm B}\ln2$ per unit cell.
This ground state is a prototypic Mott insulator. 
Assume that an electron is removed from the Mott insulator or a hole is introduced to it. There is no reason why the hole is localized in a periodic system, or the Hubbard model; the bandwidth of the hole motion is $O(W)$, i.e., the bandwidth of LHB is $O(W)$. When $U/W=+\infty$, the ground state is an insulator for $n=1$ and is a metal for $n\ne 1$. 
Next, reduce $U$. LHB is centered at the original band center $\epsilon_a$ and UHB is centered at $\epsilon_a+U$. Their bandwidths are modified by the virtual exchange processes allowing empty and double occupancies but are still $O(W)$. When $U$ is sufficiently large, the Hubbard gap cannot close.
Thus, one may argue that when $n=1$ and $U$ is sufficiently large the ground state is an insulator. However, this argument is too naive to conclude that, when $U$ is finite, the ground state can be an insulator.
According to Gutzwiller's theory \cite{Gutzwiller1,Gutzwiller2,Gutzwiller3} together with the Fermi-liquid theory, \cite{Luttinger1,Luttinger2} a narrow quasi-particle band forms near the chemical potential; the band is called the Gutzwiller band in this paper.
The ground state can only be an insulator 
if the Gutzwiller band disappears.
According to Brinkman and Rice's theory, \cite{brinkman} which is in the Gutzwiller approximation, 
the Gutzwiller band disappears when $n=1$ and $U\ge U_{\rm BR}$, where 
$U_{\rm BR}\simeq W$.
% and the specific-heat coefficient diverges as $U\rightarrow U_{\rm BR}$. 
%
In the Gutzwiller approximation, 
the ground state is an insulator when $n=1$ and $U\ge U_{\rm BR}$ and is a metal when $n\ne 1$ or $U< U_{\rm BR}$. 
The divergence of the specific-heat coefficient as $U\rightarrow U_{\rm BR}$ implies that the third law appears to be broken in the insulator. \cite{brinkman}

One may speculate based on both Hubbard's and Gutzwiller's theories that the density of states has a three-peak structure, with the Gutzwiller band between UHB and LHB. 
The approximations used in Hubbard's and Gutzwiller's theories are within the single-site approximation (SSA); they are called the Hubbard and Gutzwiller approximations, respectively. According to another SSA theory based on a slave-boson or auxiliary-particle Hubbard model, \cite{OhkawaSlave} the Gutzwiller band forms at the top of LHB when $n<1$, which implies that it forms at the bottom of UHB when $n>1$. 
The spatial dimensionality is denoted by $d$.
The SSA that considers all the single-site terms is rigorous in the limit of $d\rightarrow+\infty$,\cite{Metzner,Muller-H1,Muller-H2,Janis} but within the constrained Hilbert space where no order parameter exists; \cite{comMulti} 
this SSA is called the supreme SSA (S$^3$A).
The S$^3$A theory is mapped to a problem of self-consistently determining and solving the Anderson model, \cite{Mapping-1,Mapping-2,Mapping-3} which is an effective Hamiltonian to study the Kondo effect. 
The three-peak structure corresponds to that in the Anderson model, with the Kondo peak between two subpeaks. 
The Kondo effect is crucial for electron correlations in the Hubbard model.
The S$^3$A is also formulated as the dynamical mean-field theory\cite{georges,RevMod,kotliar,PhyToday} (DMFT) and the dynamical coherent potential approximation \cite{dcpa} (DCPA).
%\cite{comEquivalence}

It is interesting to develop $1/d$ expansion or Kondo lattice theory, in which leading order terms in $1/d$ or single-site terms are non-perturbatively considered within the S$^3$A, DMFT, or DCPA and then higher-order terms in $1/d$ or multisite terms are perturbatively considered beyond it.  
In an early Kondo lattice theory,\cite{Mapping-1,Mapping-2,Mapping-3}
which was proposed about a year earlier than the first paper of DMFT, \cite{georges}
it was assumed that 
%an {\it unperturbed} state of the perturbative theory or 
the ground state in the S$^3$A is a normal Fermi liquid. 
According to numerical theories based on DMFT, \cite{PhyToday,RevMod,kotliar,bulla} on the other hand, when $U$ is sufficiently large the ground state appears to be a highly degenerate insulator for not only $n=1$ but also $n\simeq 1$. %\cite{conNonHalfIns}
The assumption in the early Kondo lattice theory should be critically examined, because the ground state may be a highly degenerate insulator, as discussed above.
However, whether or not the ground state in the S$^3$A, DMFT, or DCPA can be a highly degenerate insulator itself is also an issue to be critically examined,
because the third law is unlikely to be broken in a physically relevant model or the Hubbard model with finite $U$. 
%even within the constrained Hilbert space, \cite{comTruGnd} 
%
It is another issue whether or not the ground state can be a highly degenerate insulator beyond the S$^3$A, DMFT, or DCPA. There is evidence that the ground state is a singlet \cite{comSinglet} even within the constrained Hilbert space, \cite{comTruGnd} as discussed below.

The Heisenberg model is derived from the half-filled Hubbard model.
%the superexchange interaction arises from the virtual exchange processes allowing empty and double occupancies in the original Hubbard model. \cite{andersonJ}
% 
Fazekas and Anderson proposed that, in the two-dimensional triangular lattice, the ground state is a spin liquid stabilized by the formation of a local singlet or a resonating valence bond (RVB) on each pair of nearest neighbors by the superexchange interaction; \cite{fazekas} this spin liquid is called an RVB spin liquid and this stabilization mechanism is called an RVB stabilization mechanism or simply an RVB mechanism. The ground state of an RVB spin liquid is a singlet. \cite{comSinglet} 
Although an RVB spin liquid is an insulator, it should be distinguished from an insulator whose ground state is highly degenerate.
In this paper, an insulator whose ground state is a singlet  \cite{comSinglet} is called a spin liquid and an insulator whose ground state is highly degenerate is only called a Mott insulator. 

The $t$-$J$ model is also derived from the Hubbard model and is reduced to the Heisenberg model when $n=1$. 
Anderson proposed that, in the two-dimensional square lattice, the ground state for $n\simeq 1$ is an exotic electron liquid,\cite{highTcRVB}
% in which spin-charge separation occurs; 
which is called an RVB electron liquid. %
The ground state within the constrained Hilbert space is certain to be a singlet \cite{comSinglet} for any $n$; it is stabilized by the RVB mechanism when $n\simeq 1$ and is a normal Fermi liquid when $n\simeq 0$.
In general, no symmetry breaking occurs when a singlet ground state is transformed into another singlet ground state as a function of a parameter such as $n$, or the adiabatic continuation holds between the two singlets.\cite{AndersonText} 
Thus, it is an issue whether the ground state for $n\simeq 1$ is really an exotic liquid or is simply a normal Fermi liquid.
The RVB stabilization is due to the Fock-type term of the superexchange interaction, either when the ground state is an RVB electron liquid \cite{Plain-vanilla} or when it is a normal Fermi liquid. \cite{phase-diagram,sc-sc}
When the RVB mechanism is considered in the Hubbard model,  
the ground state is certain to be a singlet, \cite{comSinglet} 
an exotic or normal Fermi liquid.

The Bethe ansatz solution for the Hubbard model in one dimension was obtained by Lieb and Wu. \cite{Lieb-Wu}
Two {\it chemical potentials} are defined by two processes of adding and removing a single electron:
$\mu_{+}(N)= E_{g}(N+1)- E_{g}(N)$
and
$\mu_{-}(N)= E_{g}(N)- E_{g}(N-1)$,
where $E_{g}(N)$ is the ground-state energy for the electron number $N$. Then, a gap is defined by
\begin{align}\label{EqBethe-Gap3}
\epsilon_g (N) =\mu_{+}(N)-\mu_{-}(N).
\end{align}
The number of unit cells is denoted by $L$.
According to the Bethe ansatz solution, $\epsilon_g(L)>0$ for any nonzero $U$ in the thermodynamic limit of $L\rightarrow +\infty$.
Then, Lieb and Wu argued that the half-filled ground state is an insulator for any nonzero $U$.
If there is no other solution besides the Bethe ansatz solution,\cite{comEntropy} the ground state for finite $U$ is a singlet.\cite{comSinglet} 
The insulator argued by Lieb and Wu, which is a singlet,\cite{comSinglet} is a spin liquid but is not a Mott insulator, which is highly degenerate.

The Bethe-ansatz solution is for
the canonical ensemble, in which the electron number $N$ is a constant of motion and is definitely an integer.
In the grand canonical ensemble or in the presence of an electron reservoir, on the other hand, $N$ is not a constant of motion so that the quantum-mechanically averaged electron number, which is denoted by $\left<N\right>$, is not an integer, in general. Thus,
whether the gap $\epsilon_g(N)$ is zero or nonzero is not a relevant criterion whether the ground state is a metal or an insulator. 
The compressibility is defined by
$\kappa = (1/L)(d\left<N\right>/d\mu)$,
%
%\begin{align}
%\kappa = \frac1{L}\frac{d\left<N\right>}{d\mu},
%\end{align}
%
where $\mu$ is the conventional chemical potential. Since $\left<N\right>$ is not an integer in general, what can be concluded from $\epsilon_g(L)>0$ is that the half-filled ground state is either an insulator, in which $\kappa$ is exactly zero, or a metal in which $\kappa$ is infinitesimally small but nonzero in the thermodynamic limit.\cite{comFineParticle}
%when the long range Coulomb interaction is considered, 
%
Even if $\epsilon_g(L)>0$ in the canonical ensemble, it is necessary to examine whether or not a gap opens in the single-particle excitation spectrum, which is defined in the grand canonical ensemble, of the half-filled ground state.

The main purpose of this paper is to study the ground state of the Hubbard model in the presence of an electron reservoir, but within the constrained Hilbert space where no order parameter exists.
This paper is organized as follows:
Preliminaries are given in \mbox{\S\hskip2pt\ref{SecPerliminaries}}.
The self-energy of electrons is decomposed into the single-site and multisite self-energies.
In \mbox{\S\hskip2pt\ref{SecMapping}}, it is proved that the single-site self-energy is that of a normal Fermi liquid even if the multisite self-energy is anomalous, which leads to another proof that the ground state is a normal Fermi liquid in the S$^3$A. An infinitesimally small reservoir effect is crucial in these proofs.
In \mbox{\S\hskip2pt\ref{SecHalf}}, 
the nature of the ground state is studied and
it is shown that the ground state cannot be 
%an insulator with a complete gap, nor 
a Mott insulator even beyond the S$^3$A.
In \mbox{\S\hskip2pt\ref{SecFock}}, it is shown that the normal Fermi liquid in the S$^3$A is further stabilized by the RVB mechanism. A discussion is given in \mbox{\S\hskip2pt\ref{SecDiscussion}} and a conclusion is given in \mbox{\S\hskip2pt\ref{SecConclusion}}. In Appendix\hskip1pt\ref{SecProof}, an inequality is proved for the proofs in \mbox{\S\hskip2pt\ref{SecMapping}}.
In Appendix\hskip1pt\ref{SecKL-theory}, intersite exchange interactions are reviewed for the study in \mbox{\S\hskip2pt\ref{SecFock}}.
In Appendix\hskip1pt\ref{SecDefMottInsulator}, it is argued based on the Curie-Weiss law for local moment magnetism that a Mott insulator is only possible as a high temperature phase. 

%%%%%%%%%%%%%%%%%%%%%%%%%%%%%%%%%%%%%%%%%%%%%%%%%%%%%%
%%%%%%%%%%%%%%%%%%%%%%%%%%%%%%%%%%%%%%%%%%%%%%%%%%%%%%
%%%%%%%%%%%%%%%%%%%%%%%%%%%%%%%%%%%%%%%%%%%%%%%%%%%%%%
\section{Preliminaries}
\label{SecPerliminaries}
\subsection{Electron reservoir}
%
%In general, the quantum-mechanically averaged number $\left<N\right>$ of electrons is not an integer in the grand canonical ensemble, as discussed in \S\ref{SecIntroduction}.
%In this paper, an electron reservoir is explicitly considered to include a crucial effect that is only possible when the quantum-mechanically averaged electron number $\left<N\right>$ is not an integer; it should be mentioned that a term due to the electron reservoir responsible for the crucial effect is infinitesimally small. 

The total Hamiltonian is composed of three terms:
\begin{equation}
{\cal H} = {\cal H}_{a} + {\cal H}_{b} + {\cal V}.
\end{equation}
The first term is the Hubbard model on a hypercubic lattice in $d$ dimensions, which is called an $a$ sublattice:
\begin{equation}\label{EqHubbard}
{\cal H}_a = \epsilon_a
\sum_{i\sigma}n_{i\sigma} - \frac{t}{\sqrt{d}} \sum_{\left<ij\right>\sigma} 
a_{i\sigma}^\dag a_{j\sigma}
+U \sum_{i} n_{i\uparrow} n_{i\downarrow},
\end{equation}
where $n_{i\sigma}= a_{i\sigma}^\dag a_{i\sigma}$, $\epsilon_a$ is the band center, $-t/\sqrt{d}$ is the transfer integral between nearest neighbors $\left<ij\right>$, and $U$ is the on-site repulsion;
here and in the following part of this paper, for convenience, the transfer integral is defined in such a way that it includes the dimensional factor $d$ such as $-t/\sqrt{d}$ between nearest neighbors.\cite{Metzner}
The second term represents an electron reservoir on a hypercubic lattice in $d$ dimensions, which is called a $b$ sublattice:
\begin{equation}
{\cal H}_{b} = \epsilon_b
\sum_{i\sigma}b_{i\sigma}^\dag b_{i\sigma}
- \frac{t_b}{\sqrt{d}} \sum_{\left<ij\right>\sigma} b_{i\sigma}^\dag b_{j\sigma},
\end{equation}
where $\epsilon_b$ is the band center and $-t_b/\sqrt{d}$ is the transfer integral between nearest neighbors.
The third term represents an infinitesimally small but random hybridization between the Hubbard model and the reservoir:
\begin{equation}
{\cal V} = \lambda \sum_{(ij) \in {\cal R}}\hskip-3pt\sum_{\sigma}\left[
v_{(ij)} a_{i\sigma}^\dag b_{j \sigma}
+v_{(ij)}^* b_{j\sigma}^\dag a_{i \sigma} \right] ,
\end{equation}
where $v_{(ij)}$ is a hybridization matrix between the $i$th site in the $a$ sublattice and the $j$th site in the $b$ sublattice in a set of ${\cal R}$; for convenience, an infinitesimally small but nonzero numerical constant 
\begin{align}
\lambda= \pm0^+,
\end{align}
is introduced. In order to recover the translational symmetry in an ensemble averaged system, it is assumed that
%
%\begin{equation}
$\bigl<\hskip-3pt\bigl<\hskip1pt v_{(ij)}
\hskip1pt\bigr>\hskip-3pt\bigr> = 0$ 
%\bigl<\hskip-3pt\bigl<\hskip1pt v_{(ij)}^*
%\hskip1pt\bigr>\hskip-3pt\bigr> =0 $
%\end{equation}
and 
%\begin{equation}
$\bigl<\hskip-3pt\bigl<\hskip1pt v_{(ij)} v_{(i^\prime j^\prime)}^*
\hskip1pt\bigr>\hskip-3pt\bigr> =
n_{\rm h}|v|^2 \delta_{(ij)(i^\prime j^\prime)} $, 
%\end{equation}
% 
where $\left<\hskip-2pt\left<\hskip1pt\cdots\hskip1pt\right>\hskip-2pt\right>$ represents an ensemble average over ${\cal R}$ and $n_{\rm h}$ is the density of pair of hybridization sites $(ij)$ per unit cell referring to the $a$ sublattice.
The numbers of unit cells in the $a$ and $b$ sublattices are denoted by $L$ and $L_b$. The thermodynamic limit is assumed: $L\rightarrow +\infty$ and $L_b/L\rightarrow +\infty$.

When the lattice constant of both the $a$ and $b$ sublattices is $\ell$,
the structure factor of the sublattices is given by
\begin{equation}\label{EqFormF}
f_\ell({\bm k}) = \sum_{\nu=1}^{d}\cos\left(k_{\nu}\ell\right).
\end{equation}
When $U=0$, the Green function for electrons in the Hubbard model averaged over the ensemble is given by
\begin{equation}\label{EqG0}
G_{\sigma}^{(0)}({\rm i}\varepsilon_l,{\bm k})=
\frac1{{\rm i}\varepsilon_l +\mu - \epsilon_a- E({\bm k}) 
+ \Gamma ({\rm i}\varepsilon_l) },
\end{equation}
where $\varepsilon_l=(2l+1)\pi k_{\rm B}T$, with $l$ an integer, is the fermionic energy, $\mu$ is the chemical potential, 
\begin{equation}\label{EqE(k)}
E({\bm k}) = -2\bigl(t/\sqrt{d}\bigr) f_\ell({\bm k}),
\end{equation}
and $\Gamma({\rm i}\varepsilon_l)$ is the self-energy due to scatterings from the random hybridization or a reservoir term. The bandwidth of $E({\bm k})$ is $W=O(|t|)$.
Since $\lambda= \pm0^+$, the second-order perturbation is sufficiently accurate to treat the reservoir term, so that 
\begin{equation}\label{EqGamma1}
\Gamma ({\rm i}\varepsilon_l) = 
 n_{\rm h} \lambda^2 |v|^2\frac1{L_{b}}\sum_{\bm k}
\frac1{{\rm i}\varepsilon_l +\mu - \epsilon_b - E_b({\bm k})} ,
\end{equation}
where 
%\begin{equation}
$E_b({\bm k}) = -2\bigl(t_b/\sqrt{d}\bigr) f_\ell({\bm k})$. 
%\end{equation}
%
It is assumed that 
%$\Gamma (\varepsilon+{\rm i}0)$ is continuous at $\varepsilon=0$ and
%
\begin{equation}\label{EqGamma2}
\mbox{Im} \hskip2pt \Gamma(+{\rm i}0)< 0. 
\end{equation}

The reservoir term  
ensures the quality of the grand canonical ensemble that the number of electrons in the Hubbard model, 
%\begin{equation}
$\left<N\right>= \sum_{i\sigma}\left<n_{i\sigma}\right> $,
%\end{equation}
is a continuous function of the chemical potential $\mu$.
It is assumed that in the half-filling case, in which $\left<N\right>=L$, there exists the particle-hole symmetry in the total system composed of the Hubbard model and the reservoir averaged over the ensemble.

%%%%%%%%%%%%%%%%%%%%%%%%%%%%%%%%%%%%%%%%%%%%%%%%%%%%%%
%%%%%%%%%%%%%%%%%%%%%%%%%%%%%%%%%%%%%%%%%%%%%%%%%%%%%%
%%%%%%%%%%%%%%%%%%%%%%%%%%%%%%%%%%%%%%%%%%%%%%%%%%%%%%
\subsection{Fermi surface condition}
The Anderson model is defined by
\begin{align}\label{EqAnderson}
{\cal H}_{\rm A} &=
\epsilon_d \sum_{\sigma}n_{d\sigma}
+ \sum_{{\bm k}\sigma} E_c({\bm k}) c_{{\bm k}\sigma}^\dag c_{{\bm k}\sigma}
+ U n_{d\uparrow} n_{d\downarrow}
\nonumber \\ & \qquad 
+ \frac1{\sqrt{ L_{\rm A} }} \sum_{{\bm k}\sigma} \left(
V_{\bm k}c_{{\bm k}\sigma}^\dag d_\sigma
+ V_{\bm k}^* d_\sigma^\dag c_{{\bm k}\sigma}\right) ,
\end{align} 
where $\epsilon_d$ is the level of $d$ electrons, $n_{d\sigma}=d_{\sigma}^\dag d_{\sigma}$, $E_c({\bm k})$ is the dispersion relation of conduction electrons, $V_{\bm k}$ is the hybridization matrix, and $L_{\rm A}$ is the number of unit cells; for convenience, the on-site $U$ is assumed to be the same as that in the Hubbard model. 
The Green function for $d$ electrons is given by
\begin{equation}
\tilde{G}_{\sigma}({\rm i}\varepsilon_l) =
%\frac1{\displaystyle
\left[{\rm i}\varepsilon_l+\tilde{\mu}-\epsilon_d - \tilde{\Sigma}_\sigma({\rm i}\varepsilon_l)
- \frac1{\pi}\hskip-3pt \int \hskip-3pt 
d\epsilon^\prime \frac{\Delta(\epsilon^\prime) }
{{\rm i}\varepsilon_l- \epsilon^\prime}\right]^{-1},
\end{equation}
where $\tilde{\mu}$ is the chemical potential, $\tilde{\Sigma}_\sigma({\rm i}\varepsilon_l)$ is the self-energy, and
$\Delta(\epsilon)$ is the hybridization energy defined by
\begin{equation}\label{EqDelta}
\Delta(\epsilon) =
\frac{\pi}{L_{\rm A}} \sum_{\bm k} |V_{\bm k}|^2 
\delta\bigl[\epsilon+\tilde{\mu}- E_c({\bm k})\bigr] .
\end{equation}
%.
%
The Fermi surface of conduction electrons, which is defined by $E_c({\bm k})=\tilde{\mu}$, exists provided that
\begin{equation}\label{EqFSCondition}
\Delta(0) > 0 .
\end{equation}
This condition is called the Fermi surface condition. 

The \mbox{$s$-$d$} model is another effective Hamiltonian to study the Kondo effect.
According to Yosida's perturbation theory \cite{yosida}
and Wilson's renormalization group (RNG) theory, \cite{wilsonKG} 
the ground state is a singlet or a normal Fermi liquid except when $J_{s\mbox{-}d}=0$, where $J_{s\mbox{-}d}$ is the $s$-$d$ exchange interaction. Since the \mbox{$s$-$d$} model is derived from the Anderson model, the result for the $s$-$d$ model implies that
the ground state of the Anderson model is also a normal Fermi liquid.
According to the Bethe ansatz solution for the Anderson model with nonzero and constant $\Delta(\epsilon)$, the ground state is actually a normal Fermi liquid. \cite{exact1,exact2,exact3,exact4} 
In general, the nature of the ground state depends only on relevant low-energy properties, such as $\Delta(0)$, and high-energy properties only quantitatively renormalize the ground state, as demonstrated by RNG theories for the $s$-$d$ model.\cite{wilsonKG,poorman} 
When the Fermi surface condition (\ref{EqFSCondition}) is satisfied, the ground state of the Anderson model is a normal Fermi liquid except when $U/\Delta(0)=+\infty$ and $\left<n_d\right>=1$, which corresponds to the case of $J_{s\mbox{-}d}=0$ in the \mbox{$s$-$d$} model.

%\section{Mapping to the Anderson model}
\section{Self-Consistent Single-Site Self-Energy}
\label{SecMapping}

In this section, no assumption is made for the on-site $U$, temperature $T$, and the electron density 
%\begin{align}
$n=\left<N\right>/L$.
%\end{align}
%
Within the constrained Hilbert space, in general, the Green function for electrons in the Hubbard model averaged over the ensemble is given in the wave-number representation by
\begin{equation}
G_\sigma({\rm i}\varepsilon_l,{\bm k}) =
\frac1{{\rm i}\varepsilon_l+\mu -\epsilon_a -E({\bm k}) - \Sigma_\sigma({\rm i}\varepsilon_l,{\bm k}) -\Gamma({\rm i}\varepsilon_l)},
\end{equation}
where $\Sigma_\sigma({\rm i}\varepsilon_l,{\bm k})$ is the self-energy; it is given in the site representation by
\begin{equation}
R_{ij\sigma}({\rm i}\varepsilon_l)=
\frac1{L} \sum_{\bm k} e^{i{\bm k}\cdot({\bm R}_{i}-{\bm R}_{j})}
G_{\sigma}({\rm i}\varepsilon_l, {\bm k}) ,
\end{equation}
where ${\bm R}_{i}$ is the position of the $i$th unit cell.

Consider the Feynman diagrams for the self-energy of electrons in the site representation.
Even if ensemble-averaged vertex corrections due to ${\cal V}$ appear in the diagrams, they can be ignored because they are $O(\lambda^4)$, where $\lambda=\pm0^+$.
If only lines of the on-site $U$ and the site-diagonal Green function, 
$R_{ii\sigma}({\rm i}\varepsilon_l)$,
appear in a diagram, it is a single-site diagram. 
If a line of the site-off-diagonal Green function, 
$R_{ij\sigma}({\rm i}\varepsilon_l)$
with $i\ne j$, appears, it is a multisite diagram. 
According to this classification the self-energy of electrons is decomposed into the single-site $\Sigma_\sigma({\rm i}\varepsilon_l)$ and the multisite $\Delta \Sigma_\sigma({\rm i}\varepsilon_l,{\bm k})$:
\begin{equation}
\Sigma_\sigma({\rm i}\varepsilon_l,{\bm k}) =
\Sigma_\sigma({\rm i}\varepsilon_l)
+\Delta \Sigma_\sigma({\rm i}\varepsilon_l,{\bm k}) .
\end{equation}

The same on-site $U$ appears in the the Feynman diagrams of the Hubbard and Anderson models.
If $\epsilon_d - \mu$ and $\Delta(\varepsilon)$
are determined to satisfy 
\begin{equation}\label{EqMap-G}
R_{ii\sigma}({\rm i}\varepsilon_l) = \tilde{G}_{\sigma}({\rm i}\varepsilon_l),
\end{equation}
the single-site self-energy
is given by the self-energy of the Anderson model:
\begin{equation}
\Sigma_\sigma({\rm i}\varepsilon_l)=\tilde{\Sigma}_\sigma({\rm i}\varepsilon_l).
\end{equation}
Equation~(\ref{EqMap-G}) cannot be satisfied unless
\begin{subequations}\label{EqMap-G1}
\begin{equation}\label{EqMap-G11}
\epsilon_a-\mu = \epsilon_d-\tilde{\mu}.
\end{equation}
It follows from eq.~(\ref{EqMap-G}) that
\begin{equation}\label{EqMap-G12}
\Delta(\epsilon) =
\text{Im}\left[
\tilde{\Sigma}_\sigma(\epsilon+{\rm i}0)
+ R_{ii\sigma}^{-1}(\epsilon+{\rm i}0)
\right].
\end{equation}
\end{subequations}
Equation~(\ref{EqMap-G}) or (\ref{EqMap-G1}) is the condition that the calculation of the single-site self-energy is mapped to a problem of self-consistently determining and solving the Anderson model;   in this paper, eq.~(\ref{EqMap-G}) or (\ref{EqMap-G1}) is called the mapping condition to the Anderson model.
It should be noted that $\Delta(\epsilon)$ or the mapped Anderson model itself may depend on $T$.

The mapping condition (\ref{EqMap-G12}) is iteratively treated to obtain the eventual self-consistent $\Delta(\epsilon)$; not only $\Delta(\epsilon)$ but also the single-site $\tilde{\Sigma}_\sigma(\epsilon)$ and the multisite $\Delta\Sigma_\sigma(\epsilon + {\rm i}0,{\bm k})$ should be self-consistently determined with each other to satisfy the mapping condition (\ref{EqMap-G12}). 
It is proved in Appendix\hskip1pt\ref{SecProof} that 
\begin{equation}\label{EqSelfDelta}
\Delta(\epsilon) \ge
-\text{Im} \Gamma (\epsilon + {\rm i}0) .
\end{equation}
According to eqs.~(\ref{EqGamma2}) and (\ref{EqSelfDelta}), 
the Fermi surface condition (\ref{EqFSCondition}), $\Delta(0)>0$, is satisfied at each step of the iterative process, so that the ground state of the self-consistently determined or mapped Anderson model should be a normal Fermi liquid even if the multisite self-energy is anomalous or the ground state of the Hubbard model is a non-normal Fermi liquid.
Thus, it is proved that the single-site self-energy for the ground state is that of the normal Fermi liquid in the Anderson model. Then, it is also proved that, when the multisite self-energy is ignored in the S$^3$A, the ground state of the Hubbard model is a normal Fermi liquid.

Consider the mapped Anderson model.
Then, introduce an infinitesimally small Zeeman energy $h=g\mu_{\rm B}H$.
Since the ground state of the Anderson model is a normal Fermi liquid, the self-energy of the Anderson model is expanded in such a way that, for example, at $T=0$~K, 
\begin{align}\label{EqSelfExpansion}
\tilde{\Sigma}_\sigma(\epsilon+{\rm i}0) &= 
\tilde{\Sigma}_0 + (1 - \tilde{\phi}_{\rm e})\epsilon 
- (\alpha+ i \gamma) \epsilon^2/k_{\rm B}T_{\rm K} 
\nonumber \\ & \quad
- (1 - \tilde{\phi}_{\rm s} )\frac1{2}\sigma h 
+ O\left[\epsilon^3/(k_{\rm B}T_{\rm K})^2 \right],
\end{align}
where $\tilde{\Sigma}_0$, $\tilde{\phi}_{\rm e}>0$, $\tilde{\phi}_{\rm s}>0$, $\alpha$, and $\gamma>0$ are all real; $T_{\rm K}$ or $k_{\rm B}T_{\rm K}$ is the Kondo temperature, which is the energy scale of local quantum spin fluctuations in not only the Anderson model but also the Hubbard model.
According to the proofs made above in this section,  
the expansion (\ref{EqSelfExpansion}) is relevant even if the ground state of the Hubbard model is a non-normal Fermi liquid, and 
$k_{\rm B}T_{\rm K}$ is definitely nonzero although it may be extremely or infinitesimally small.

%%%%%%%%%%%%%%%%%%%%%%%%%%%%%%%%%%%%%%%%%%%%
%%%%%%%%%%%%%%%%%%%%%%%%%%%%%%%%%%%%%%%%%%%%
%\section{Half-filled ground state}
\section{Nature of the Ground State}
\label{SecHalf}
In this section, the nature of 
the ground state in the weak and strong-coupling regimes, 
%\begin{align}
$0 < U/|t| < +\infty$, 
%\end{align}
%
is studied under only the assumption that the multisite self-energy $\Delta\Sigma_\sigma(\epsilon+{\rm i}0,{\bm k})$ is analytic in the upper half plane; it may be convergent or divergent on the real axis.

The single-particle excitation spectrum is given by
\begin{equation}\label{EqRho2}
\rho(\epsilon) = -\frac1{\pi}{\rm Im}R_{ii\sigma}(\epsilon+{\rm i}0)
= -\frac1{\pi}{\rm Im}\tilde{G}_\sigma(\epsilon+{\rm i}0),
\end{equation}
where the mapping condition (\ref{EqMap-G}) is made use of.
The density of states in the Hubbard model is the same as that in the mapped Anderson model.
It follows from eqs.~(\ref{EqSelfExpansion}) and (\ref{EqRho2}) that 
\begin{subequations}\label{EqRho30}
\begin{align}\label{EqRho31}
\rho(\epsilon) &= 
-\frac1{\pi L}\sum_{\bm k}{\rm Im}
\frac1{M(\epsilon)- E({\bm k})
- \Delta\Sigma_\sigma(\epsilon+{\rm i}0,{\bm k})}
\\ \label{EqRho32} &=
-\frac1{\pi}{\rm Im}
\left[
M(\epsilon)
- \frac1{\pi}\hskip-3pt \int \hskip-3pt 
d\epsilon^\prime \frac{\Delta(\epsilon^\prime) }
{\epsilon- \epsilon^\prime + {\rm i}0} \right]^{-1},
\end{align}
\end{subequations}
where $E({\bm k})$ is defined by Eq.~(\ref{EqE(k)});
when $|\epsilon|\ll k_{\rm B}T_{\rm K}$, 
% $M(\epsilon)$ is defined by
%
\begin{align}\label{EqSelfExpansion2}
M(\epsilon) &= 
%\epsilon + \mu -\epsilon_a -\tilde{\Sigma}_\sigma( \epsilon+{\rm i}0)
%%\nonumber \\ &=
\Delta\mu + \tilde{\phi}_{\rm e}\epsilon 
+ (\alpha+ i \gamma) \epsilon^2/k_{\rm B}T_{\rm K}
%
%\nonumber \\ & \qquad
+ O\left[\epsilon^3/(k_{\rm B}T_{\rm K})^2 \right],
\end{align}
where
\begin{equation}\label{EqDeltaMu}
\Delta\mu=\mu-\epsilon_a -\tilde{\Sigma}_0. 
\end{equation}

The half-filling case $(n=1)$ is first studied.
Since the particle-hole symmetry exists in not only the Hubbard model but also the Anderson model, it follows that
\begin{equation}\label{EqPH1}
\tilde{\Sigma}_0 = \frac1{2}U= \tilde{\mu} -\epsilon_d =\mu -\epsilon_a,
\end{equation}
\begin{equation}\label{EqPH2}
\Delta\mu=0, \quad \alpha=0, \quad 
\Delta(\epsilon) = \Delta(-\epsilon),
\end{equation}
and
\begin{equation}
\left[\int \hskip-3pt 
d\epsilon^\prime \frac{\Delta(\epsilon^\prime) }
{\epsilon- \epsilon^\prime + {\rm i}0}\right]_{\epsilon=0} =- i \pi \Delta(0).
\end{equation}
Then, it follows from eq.~(\ref{EqRho32}) that 
\begin{equation}\label{EqRhoA}
\rho(0) = 1/\left[\pi \Delta(0)\right]. 
\end{equation}
{\em When $\rho(0)$ is vanishing or the ground state continuously approaches to an insulating state, $\Delta(0)$ is diverging such that $\Delta(0)/|t|>0$ and $\Delta(0)/|t|\rightarrow +\infty$. Even if $\rho(0)|t|\rightarrow 0^+$, the Fermi surface condition is definitely satisfied.}

%the multisite self-energy vanishes such as

In the S$^3$A, $\Delta\Sigma_\sigma(\epsilon+{\rm i}0,{\bm k})=0$. The ground state is definitely a normal Fermi liquid and no gap opens in $\rho(\epsilon)$.
Since $\Delta\mu=0$, as shown in eq.~(\ref{EqPH2}),
the Fermi surface is a hypersurface defined by
\begin{equation}\label{EqHalfFilledFS}
E({\bm k})=0.
\end{equation}
It follows from eq.~(\ref{EqRho31}) that
\begin{equation}\label{EqRho0}
\rho(0) = \frac1{L}\sum_{\bm k} \delta\bigl[E({\bm k})\bigr]. 
\end{equation}
Thus, $\rho(0)$ does not depend on $U$;
$\Delta (0)$ does not depend on $U$ either
according to eqs.~(\ref{EqRhoA}) and (\ref{EqRho0}).

Beyond the S$^3$A, $\Delta\Sigma_\sigma(\epsilon+{\rm i}0,{\bm k})$ can be nonzero. 
Assume that it is continuous and finite at $\epsilon=0$ for any ${\bm k}$.
According to the particle-hole symmetry, it follows that
\begin{equation}
E({\bm k}) +
{\rm Re}\Delta\Sigma_\sigma(+{\rm i}0,{\bm k}) =0,
\end{equation}
for any ${\bm k}$ defined by eq.~(\ref{EqHalfFilledFS}).
Then, $\rho(0)>0$, i.e., no gap can open in $\rho(\epsilon)$. The ground state is a metal; it may be a normal or anomalous Fermi liquid.

Since $\Delta\Sigma_\sigma(\epsilon + {\rm i}0,{\bm k})$ is analytic, it can only diverge at a pole or an end-point of a cut; it cannot diverge on a line or an area.
Then, assume that $\Delta\Sigma_\sigma(\epsilon+ {\rm i}0,{\bm k})$ is diverging as $\epsilon\rightarrow\pm0$ on the real axis: 
\begin{equation}\label{EqDiv}
\lim_{\epsilon\rightarrow \pm0}\left|
\Delta\Sigma_\sigma(\epsilon+ {\rm i}0,{\bm k}) \right| = +\infty. \end{equation}
If the divergence (\ref{EqDiv}) occurs for a part of ${\bm k}$'s defined by eq.~(\ref{EqHalfFilledFS}), $\rho(0)>0$ and the ground state is a metal.
If the divergence (\ref{EqDiv}) occurs at least for all the ${\bm k}$'s defined by eq.~(\ref{EqHalfFilledFS}), $\rho(0)=0$.
Since $i\gamma \epsilon^2/(k_{\rm B}T_{\rm K})$ is nonzero and $\Delta\Sigma_\sigma(\epsilon+ {\rm i}0,{\bm k})$ is finite for $\epsilon\ne0$, $\rho(\epsilon)>0$ for at least $0<|\epsilon|\ll k_{\rm B}T_{\rm K}$. Then, a pseudogap opens and the ground state is a gapless semiconductor.
Even if the divergence (\ref{EqDiv}) occurs for any ${\bm k}$, the ground state cannot be an insulator with a complete gap.

It is straightforward to extend the above analysis to nonhalf fillings $(n\ne 1)$. 
If no multisite self-energy is considered in the S$^3$A,
the ground state is a normal Fermi liquid.
If the multisite self-energy $\Delta\Sigma_\sigma(\epsilon+ {\rm i}0,{\bm k})$ is continuous and finite at $\epsilon=0$, the ground state is a metal.
Even if the divergence (\ref{EqDiv}) occurs for any ${\bm k}$, the ground state cannot be an insulator with a complete gap but can only be a gapless semiconductor.

\section{RVB Stabilization Mechanism}
\label{SecFock}
In this section, the ground state in the strong-coupling regime,
$1\ll U/|t| <+\infty$, is studied. 
According to previous studies for the $t$-$J$ model, \cite{phase-diagram,sc-sc} the Fermi liquid in S$^3$A is further stabilized by the Fock-type term of the superexchange interaction. 
In this section, 
the study for the $t$-$J$ model is extended to the Hubbard model.

In the second-order perturbation in $t$, the superexchange interaction arises from the virtual exchange processes allowing empty and double occupancies.\cite{andersonJ} 
In field theory, it arises from the virtual exchange of a pair excitation of electrons across the Hubbard gap: \cite{three-exchange,itinerant-ferro,sup-exchange}
\begin{equation}
J_s({\bm q}) = 2 (J/d) f_\ell({\bm q}),
\end{equation}
where $f_\ell({\bm q})$ is define by Eq.~(\ref{EqFormF}) and
\begin{align}
J/d = -4 \alpha t^2/dU,
\end{align}
where $\alpha$ is a numerical constant; $\alpha=1$ when the bandwidths of UHB and LHB are ignored, but $0<\alpha<1$ when they are considered. \cite{exchange-reduction} 
An effective three-point single-site vertex function in spin channels is given by the expansion coefficient $\tilde{\phi}_{\rm s}$ of the single-site self-energy, as shown in \mbox{Appendix\hskip1pt\ref{SecKL-theory}}.

When the expansion (\ref{EqSelfExpansion}) is used, the coherent part of the Green function is given by
\begin{equation}\label{EqG-RVB}
G_\sigma({\rm i}\varepsilon_l,{\bm k}) =
\frac1{\tilde{\phi}_{\rm e}\epsilon + \Delta\mu - E({\bm k}) 
- \Delta\Sigma_\sigma({\rm i}\varepsilon_l,{\bm k}) } + \cdots,
\end{equation}
where $\Delta\mu$ is defined by eq.~(\ref{EqDeltaMu}); $(\alpha+i\gamma) \epsilon^2/(k_{\rm B}T_{\rm K})$ is ignored.
When only the coherent part is considered, the Fock-type term is determined from
\begin{align}\label{EqSelf-1}
\Delta\Sigma_\sigma({\rm i}\varepsilon_l,{\bm k}) &= 
 k_{\rm B} T \frac1{L} 
\sum_{\varepsilon_l{\bm p}}e^{i \varepsilon_{l}0^+}
\frac{3}{4} \tilde{\phi}_{\rm s}^2 J_s({\bm k}- {\bm p}) 
\nonumber \\ & \times
\frac1
{i\tilde{\phi}_{\rm e}
 \varepsilon_{l} + \Delta\mu - E({\bm p})
 - \Delta\Sigma_\sigma({\rm i}\varepsilon_l,{\bm p})} .
\end{align}
%
%The factor 3 appears because of three spin channels.
It follows from this self-consistent equation that
\begin{equation}\label{EqDeltaSigma}
\Delta \Sigma_\sigma (\epsilon+{\rm i}0, {\bm k}) =
\frac{1}{4}\tilde{\phi}_{\rm e} c_J (J/d)
f_\ell({\bm k}) , 
\end{equation}
where
\begin{equation}\label{EqXi}
c_J =
\frac{3}{d}\left(\tilde{\phi}_{\rm s}/\tilde{\phi}_{\rm e}\right)^2
\hskip-3pt
\frac1{L}\sum_{{\bm k}} \theta \bigl[-\xi({\bm k})/|t|\bigr]
 f_\ell({\bm k}) ,
\end{equation}
where $\theta(\epsilon)$ is defined by
%\begin{align}
$\theta(\epsilon\ge  0) =1$ and $\theta(\epsilon< 0) =0$; 
%\end{align}
and
$\xi({\bm k})$ is the pole of eq.~(\ref{EqG-RVB}), so that
\begin{align}
\xi({\bm k}) &= 
%\frac1{\tilde{\phi}_{\rm e}}\left[E({\bm k})+ 
%\Delta\Sigma_\sigma(\epsilon+{\rm i}0,{\bm k})\right]
%\nonumber \\ &=
\frac1{\tilde{\phi}_{\rm e}}\left[ -2 (t^*/\sqrt{d}) f_\ell({\bm k})
- \Delta\mu \right],
\end{align}
where
\begin{equation}
t^* = t - 
(1/8)\tilde{\phi}_{\rm e} c_J (J/\sqrt{d}) .
\end{equation}
The Fock-type term is of higher order in $1/d$.

Since the Fock-type term (\ref{EqDeltaSigma}) does not depend on energy $\epsilon$ so that it is normal, the ground state in this approximation is a normal Fermi liquid. Then, $\xi({\bm k})$ is simply the dispersion relation of quasi-particle in the normal Fermi liquid.
According to the Fermi surface sum rule,\cite{Luttinger1,Luttinger2} $\Delta\mu$ can be determined by 
\begin{equation}
n =\frac{2}{L}\sum_{\bm k} \theta\left[-\xi({\bm k})/|t|\right].
\end{equation}
%
%where $n$ is the number of electrons per unit cell.
According to the Fermi-liquid theory, \cite{Luttinger1,Luttinger2}
the low-temperature specific heat is proportional to $T$ such as 
%\begin{align}
$C=\gamma_C T+ \cdots$  
%\end{align}
%
and the specific coefficient is given by 
\begin{equation}\label{EqGammaC}
\gamma_C = \frac{2}{3}\pi^2 k_{\rm B}^2 \tilde{\phi}_{\rm e}\rho(0),
\end{equation}
where 
\begin{equation}\label{EqRhoFock}
\rho(0) = \frac1{L}\sum_{\bm k}
\delta\!\left[ - 2(t^*/\sqrt{d})f_\ell({\bm k})- \Delta\mu\right] . 
\end{equation}
Since $J/|t|<0$, the Fock-type term enhances the bandwidth of quasi-particles, so that the Fermi-liquid ground state in the S$^3$A is further stabilized by the RVB mechanism. However, $\rho(0)$ is reduced by the RVB mechanism; $\Delta(0)$ is enhanced according to eq.~(\ref{EqRhoA})

Low-energy local quantum spin fluctuations, whose energy scale is $k_{\rm B}T_{\rm K}$, are renormalized by the superexchange interaction or high-energy intersite spin fluctuations, whose energy scale is $O(U)$. In such a situation, the low-energy scale $k_{\rm B}T_{\rm K}$ is simply proportional to the renormalized bandwidth of quasi-particles, i.e.,  $k_{\rm B}T_{\rm K} \propto |t^*|$. In particular, when $\tilde{\phi}_{\rm e} \gg 1$ or $\tilde{\phi}_{\rm e} \rightarrow +\infty$, 
\begin{align}
k_{\rm B}T_{\rm K} \propto |J|.
\end{align}
When $U$ is finite, $k_{\rm B}T_{\rm K}$ is definitely nonzero in this approximation, even for the just half-filled case.

Consider the half-filled case and assume that $d$ is finite. 
In the limit of $U/|t|\rightarrow +\infty$ with $t^2/U$ kept constant, the Hubbard model is reduced to the Heisenberg model with the superexchange interaction constant
\begin{align}
J/d=-4 t^2/(dU),
\end{align}
between nearest neighbors; this limit is called the Heisenberg limit.
The probabilities of empty and double occupancies in the Hubbard model are $O(t^2/U^2)$, so that those in the mapped Anderson model are also $O(t^2/U^2)$.
In the Anderson model, the expansion coefficient $\tilde{\phi}_{\rm e}$ of the self-energy is inversely proportional to the probabilities, so that 
%\begin{align}
$\tilde{\phi}_{\rm e}=O(U^2/t^2)$.
%\end{align}
In the Heisenberg limit, it follows that $t/t^*\rightarrow 0$ or
\begin{equation}\label{EqRF}
\frac{t}{(1/8)\tilde{\phi}_{\rm e} c_J (J/\sqrt{d})} \rightarrow 0,
\end{equation}
which means that the normal Fermi liquid is totally stabilized by the Fock-type term. It follows that 
\begin{align}
\rho(0)|t|\rightarrow 0, \quad
\gamma_C\propto\sqrt{d}/|J|.
\end{align} 
Note that the Fermi surface condition is definitely satisfied even in the limit of $\rho(0)|t|\rightarrow 0$, as shown in \mbox{\S\hskip2pt\ref{SecHalf}}.
Since $\rho(0)$ is vanishing but the $T$-linear specific-heat coefficient $\gamma_C$ is nonzero and finite, a normal Fermi liquid in the Heisenberg limit is frustrated as much as an RVB spin liquid in the Heisenberg model. \cite{fazekas} 
%
%Since the Fock-type term is normal, as studied in \mbox{\S\hskip2pt\ref{SecFock}}, an RVB electron liquid \cite{highTcRVB} is simply a normal Fermi liquid; but it is highly frustrated. 
%The study in \S\hskip2pt\ref{SecFock} implies that the adiabatic continuation \cite{AndersonText} holds among an RVB spin liquid,\cite{fazekas} an RVB electron liquid,\cite{highTcRVB} and a normal Fermi liquid, as it holds between a spin liquid in the $s$-$d$ model and an electron liquid in the Anderson model, either of which is a singlet or a normal Fermi liquid.
%An RVB spin liquid, an RVB electron liquid, and a normal Fermi liquid must be renormalized by higher-order multisite terms than the Fock-type term; but  
%Thus, it is plausible that the natures of the three types of spin or electron liquids in the Heisenberg and Hubbard models are the same as each other.
It is plausible that the Fermi liquid in the Heisenberg limit is simply an RVB spin liquid in the Heisenberg model or the adiabatic continuation \cite{AndersonText} holds between the two liquids, at least in the approximation of this section, where only the Fock-type term is considered beyond the S$^3$A.

\section{Discussion}
\label{SecDiscussion}
Suppose that the infinitesimally small reservoir term exactly vanishes such that $\Gamma(\epsilon+{\rm i}0)=0$. \cite{comAssumption}
In this case, the Fermi surface condition (\ref{EqFSCondition}) may or may not be satisfied so that the ground state of the mapped Anderson model may or may not be a singlet.
 If the ground state of the Anderson model is a singlet, the ground state of the Hubbard model is a normal Fermi liquid in the S$^3$A, as studied in \S\hskip2pt\ref{SecMapping}. Then, if the ground state of the Hubbard model is not a normal Fermi liquid in the S$^3$A, the ground state of the Anderson model is not a singlet. Thus, if a non-normal Fermi-liquid ground state is possible 
in the S$^3$A, it should be highly degenerate.
This is the reason why only an insulator whose ground state is highly degenerate is called a Mott insulator in this paper.

In the so called Hubbard I approximation, \cite{Hubbard1} for example, when $n=1$ the self-energy is given by
\begin{align}\label{EqHubbardI}
\Sigma_\sigma(\epsilon+{\rm i}0,{\bm k}) = \frac{U}{2} + \frac{U^2}{4(\epsilon+{\rm i}0)},
\end{align}
which has a pole at $\epsilon=-{\rm i}0$, so that the ground state is an insulator with a complete gap.  In general,
the ground state can be an insulator with a complete gap only if the self-energy $\Sigma_\sigma(\epsilon+{\rm i}0,{\bm k})$ has a pole at $\epsilon=-{\rm i}0$;
%, i.e., an anomalous self-energy with a pole at $\epsilon=-{\rm i}0$ is only a possible route to an insulating ground state with a complete gap;
 the insulating ground state is highly degenerate, as discussed above, so that it is a Mott insulator.
As proved in \mbox{\S\hskip2pt\ref{SecMapping}}, however, any self-energy with a pole at $\epsilon=-{\rm i}0$ cannot be a self-consistent S$^3$A self-energy for satisfying the mapping condition (\ref{EqMap-G12}) if once the reservoir term $\Gamma(\epsilon+{\rm i}0)$ is introduced.
In the S$^3$A, the ground state cannot be a Mott insulator if once the reservoir term is considered.

Since the formulations of the S$^3$A, DMFT, and DCPA themselves are exactly equivalent to each other, it is surprising that numerical theories based on DMFT\cite{PhyToday,RevMod,bulla} are inconsistent with  the proof in \S\hskip1pt\ref{SecMapping}.  
For example, Bulla studied the half-filled $(n=1)$ ground state within the constrained Hilbert space by Wilson's numerical RNG method based on DMFT. \cite{bulla}
According to this numerical RNG theory, the ground state can be a Mott insulator for finite $U$, %\cite{comBulla}
which is inconsistent with the proof in \mbox{\S\hskip1pt\ref{SecMapping}} that ground state is a normal Fermi liquid for any finite $U$.
% 
%Even if a Mott insulator is possible when no reservoir term is considered, if once the reservoir term $\Gamma(\epsilon+{\rm i}0)$ is considered the Mott insulator is definitely unstable against a normal Fermi liquid in which $\rho(0)$ cannot depend on $U$ because the constancy of $\rho(0)$ is proved in \S\hskip1pt\ref{SecHalf}. 
%
A possible explanation for the inconsistency is that the reservoir term, which is infinitesimally small, is considered in the proof but no reservoir term is considered in the numerical RNG theory.
Another possible explanation for the inconsistency is that
numerical theories for $T=0$~K and $T>0$~K cannot treat a normal Fermi liquid with extremely or infinitesimally small $k_{\rm B}T_{\rm K}$ because of limited numerical accuracy and nonzero temperature. When the proof in \S\hskip1pt\ref{SecMapping} is taken into account, the numerical RNG theory \cite{bulla} for $T=0$~K simply means that, when $U\gtrsim W$ and $n= 1$, $k_{\rm B}T_{\rm K}$ is extremely or infinitesimally small in the S$^3$A, DMFT, or DCPA.
%
%When numerical theories developed for $T>0$~K are concerned, 
% 
On the other hand, a Mott insulator is only possible as a high temperature phase, as discussed in \mbox{Appendix\hskip1pt\ref{SecDefMottInsulator}}.
When $T_{\rm K}$ is extremely or infinitesimally small, numerical theories for $T>0$~K cannot treat a crossover between a normal Fermi liquid at $T\ll T_{\rm K}$ and a Mott insulator at $T\gg T_{\rm K}$. 
If an {\it insulating} phase appears at a low temperature $T>0$~K, the appearance means that $T_{\rm K}\ll T$ or $k_{\rm B}T_{\rm K}$ is extremely or infinitesimally small in the insulating phase.

In this paper, an unrealistic electron reservoir is considered in order to recover the translational symmetry in an ensemble averaged system, which simplifies the formulation as the periodic boundary condition, which is also unrealistic, simplifies the formulation.
It is desirable to consider a more realistic electron reservoir.
In general, a highly degenerate ground state is unstable against an infinitesimally small perturbation, so that the third law of thermodynamics holds in a physically relevant model.
Thus, it is unlikely that there is an S$^3$A solution in which the third law is broken, even when a more realistic electron reservoir is considered or no electron reservoir is considered. 

Even if a Mott insulator is possible in the S$^3$A with $\Gamma(\epsilon+{\rm i}0)=0$ assumed, it is unstable if once the RVB mechanism is considered beyond the S$^3$A, as discussed below. 
Single-site properties in the S$^3$A such as  $\Delta(\varepsilon)$, which is the hybridization energy of the Anderson model, $\tilde{\Sigma}_0$, $\tilde{\phi}_{\rm e}$, and $\tilde{\phi}_s$, which are expansion coefficients of the single-site self-energy, and so forth  should be self-consistently calculated with the Fock-type term of the superexchange interaction.\cite{comBeyond}
First, we assume that, even when $\Gamma(\epsilon+{\rm i}0)=0$, $\tilde{\phi}_{\rm e}$ is finite beyond the S$^3$A, 
and then we examine the relevance of the assumption of finite $\tilde{\phi}_{\rm e}$. The examination is exactly in parallel with the study in \mbox{\S\hskip2pt\ref{SecFock}}. 
Provided that $\tilde{\phi}_{\rm e}$ is finite, even if $\tilde{\phi}_{\rm e}$ is extremely large, the Kondo temperature $k_{\rm B}T_{\rm K}$ is renormalized by the RVB mechanism, so that $k_{\rm B}T_{\rm K}>O(|J|)$ or $k_{\rm B}T_{\rm K}=O(|J|)$ even if 
$\Gamma(\epsilon+i0)=0$ is assumed.
The nonzero $k_{\rm B}T_{\rm K}$ means that,  when only the Fock-type term is considered beyond the S$^3$A, a normal Fermi liquid, which is characterized by $\tilde{\phi}_{\rm e}<+\infty$, is stable against a Mott insulator, which is characterized by $\tilde{\phi}_{\rm e}=+\infty$ or the self-energy with a pole at $\epsilon=-i{\rm }0$. Thus, the assumption of finite $\tilde{\phi}_{\rm e}$ must be relevant. 
Since $k_{\rm B}T_{\rm K}>O(|J|)$ or $k_{\rm B}T_{\rm K}=O(|J|)$   beyond the S$^3$A, whether $k_{\rm B}T_{\rm K}$ is nonzero, extremely small, infinitesimally small, or exactly zero within the S$^3$A is never any crucial issue.
Since the ground state cannot be a Mott insulator beyond the S$^3$A, whether or not the ground state can be a Mott insulator within the S$^3$A is never any crucial issue either.

It is interesting to study if higher-order multisite terms than the Fock-type term can be so anomalous that the ground state is a gapless semiconductor.
The divergence of the self-energy at the chemical potential implies that the ground state is highly or infinitely degenerate. Thus, the third law of thermodynamics appears to be incompatible with the divergent self-energy. Therefore, the ground state within the constrained Hilbert space is unlikely to be a gapless semiconductor in one dimension and higher dimensions. 

It is straightforward to show that an anomalous term proportional to $\epsilon \ln\left|\epsilon\right|$ exists in the multisite self-energy in one dimension; thus, the ground state in one dimension is the so called Tomonaga-Luttinger liquid, which is not a normal Fermi liquid. Since the anomalous term is continuous and finite at $\epsilon=0$, no gap opens.
If the truth is that the self-energy is continuous and finite not only for $n\ne 1$ but also $n=1$, the ground state within the constrained Hilbert space is a metal in one dimension and higher dimensions, even beyond the S$^3$A.

%It is plausible that the adiabatic continuation \cite{AndersonText} holds between a Tomonaga-Luttinger spin liquid in the Heisenberg model and a Tmonaga-Luttinger liquid in the Hubbard model.

Suppose that an electron or hole is added onto the chemical potential in the half-filled Hubbard model. When $N$ is not a constant of motion,
% in the presence of an electron reservoir, 
the quantum-mechanically averaged number $\left<N\right>$ is not an integer, in general. If a quasi-particle band exists near the chemical potential, a quantum process is possible in which only a small or infinitesimally small fraction of the added electron or hole enters the quasi-particle band  and almost the entire component enters the reservoir; thus, no gap opens in the single-particle excitation spectrum. It is trivial that if no quasi-particle band exists a gap opens in the spectrum. 
When $N$ is restricted within integers, on the other hand, the quantum process discussed above is impossible.
The entire component of the added electron or hole remains in the Hubbard model. It is trivial that, when $U\gg W$, the gap $\epsilon_g(L)$ defined by eq.~(\ref{EqBethe-Gap3}) is nonzero such that $\epsilon_g(L) =U-O(W)>0$ in one dimension and higher dimensions. 
%, which is defined by eq.~(\ref{EqBethe-Gap3}),
When $N=L$ and $U\gg W$, a gap as large as $U$ inevitably opens in the spectrum of adding a single electron or hole, even if a quasi-particle band exists near the chemical potential.

Observable physical properties or {\it observables} of an electron liquid are directly related with the pair excitation spectrum, which is bosonic. 
Although observables are not directly related with the single-particle excitation spectrum $\rho(\epsilon)$, which is fermionic,
certain observables can be described by $\rho(\epsilon)$ according to the Fermi-liquid theory in the grand canonical ensemble, \cite{Luttinger1,Luttinger2} as shown in eq.~(\ref{EqGammaC}).
Since the virtual exchange processes of allowing empty and double occupancies are possible, a characteristic energy of low-energy pair excitations is $O(|J|)$ in the strong-coupling regime.
Then, it is quite reasonable that the energy scale of $O(|J|)$ appears in $\rho(\epsilon)$. 
Since it is unlikely that observables in the canonical ensemble
are different from those in the grand canonical ensemble, the gap $\epsilon_g(L)$ in the canonical ensemble is not directly related with observables. 
It should be examined how the gap $\epsilon_g(L)$ in the canonical ensemble are related with observables or what physical significance $\epsilon_g(L)$ has.

The relevance of the assumption in the early Kondo lattice theory \cite{Mapping-1,Mapping-2,Mapping-3} has been confirmed in this paper; thus,
the Kondo lattice theory is simply a perturbative theory starting from a normal Fermi liquid constructed in the S$^3$A to study a normal or anomalous Fermi liquid and an ordered state in the whole Hilbert space.
Not only the further stabilization of the normal Fermi liquid in the S$^3$A by the RVB mechanism,\cite{phase-diagram,sc-sc} which is also studied in \S\hskip2pt\ref{SecFock} of this paper, but also various anomalous Fermi-liquid properties can be explained by the Kondo lattice theory: paramagnon effect,\cite{Mapping-1} metamagnetic transition or crossover, \cite{satoh} and so forth. 
The Curie-Weiss (CW) law for itinerant magnetism is also an anomalous Fermi-liquid property;
in a paramagnetic phase of an itinerant electron magnet, 
the static susceptibility $\chi_s(0,{\bm q})$ obeys the CW law for only ${\bm q}$ near an ordering wave number, but not for all ${\bm q}$.  
In previous papers,\cite{three-exchange,CW1} the CW law for itinerant magnetism has already been studied by the Kondo lattice theory.  
Many anomalous Fermi-liquid properties have been observed in the cuprate oxide superconductor at $T>T_c$, where $T_c$ is superconducting critical temperature:
resistivity almost linear in $T$, a pseudogap, and so forth.
According to ref.\hskip2pt\citeonline{moriya-rho}, resistivity can increase almost linearly in $T$ because of critical antiferromagnetic fluctuations in two or highly anisotropic quasi-two dimensions. According to ref.\hskip2pt\citeonline{ps-gap2}, a pseudogap can open because of critical superconducting fluctuations in two or highly anisotropic quasi-two dimensions. Such anomalous Fermi-liquid properties at $T>T_c$ can also be treated by the Kondo lattice theory when effects of antiferromagnetic or superconducting fluctuations are included in the multisite self-energy. 

An ordered state can also be studied by the Kondo lattice theory.
% when an infinitesimally small symmetry breaking field corresponding to the ordered state is assumed. 
In ref.\hskip2pt\citeonline{itinerant-ferro}, a theory of itinerant electron ferromagnetism has already been developed based on a multiband Hubbard model, in which the superexchange interaction is ferromagnetic because of the Hund coupling.
An early Fermi-liquid theory \cite{highTc2} of high-temperature superconductivity, which was proposed just after the discovery of the cuprate oxide superconductor,\cite{bednortz} is consistent with the theory in this paper, so that it can be regarded as one of the simplest version of Kondo lattice theories of superconductivity.
In a recent paper,\cite{sc-sc} a theory of strong-coupling superconductivity has been developed to consider strong-coupling effects in the cuprate oxide superconductor.

\section{Conclusion}
\label{SecConclusion}
%
%The ground state of the Hubbard model has been studied  within the constrained Hilbert space where no order parameter exists in the grand canonical ensemble.
In the canonical ensemble, the number of electrons is a constant of motion and is definitely an integer.
In the grand canonical ensemble, on the other hand, the number of electrons within the Hubbard model is not a constant of motion because of the existence of an electron reservoir, so that the quantum-mechanically averaged electron number is, in general, not an integer.
Because of this difference between the two ensembles, 
the single-particle excitation spectrum in the grand canonical ensemble, which is the spectrum of adding an infinitesimally small number of electrons or holes, can be different from the spectrum of adding a single electron or hole in the canonical ensemble.
According to the Fermi-liquid theory, the spectrum in the grand canonical ensemble is related to observable physical properties such as the specific-heat coefficient; therefore, the spectrum in the canonical ensemble is, in general, not directly related to observable physical properties, so that whether or not a gap opens in the spectrum in the canonical ensemble is not a relevant criterion whether the ground state is a metal or an insulator.
Thus, the single-particle excitation spectrum in the grand canonical ensemble has been studied to elucidate the nature of the ground state of the Hubbard model or determine whether the ground state of the Hubbard model is a metal or an insulator, but within the constrained Hilbert space where no order parameter exists.
%rather than the spectrum of adding a single electron or hole in the canonical ensemble.

Within the constrained Hilbert space,
the self-energy of electrons is decomposed into the single-site and multisite self-energies, in general.
It has been proved that the single-site self-energy for the ground state is that of a normal Fermi liquid even if the multisite self-energy is anomalous. 
If the multisite self-energy is continuous and finite at the chemical potential, no gap opens in the single-particle excitation spectrum; thus, the ground state is a metal, a normal or anomalous Fermi liquid. 
If the multisite self-energy is so anomalous that it is divergent at the chemical potential, only a pseudogap can open such that the ground state can be merely a gapless semiconductor; 
however, it is unlikely that the ground state is a gapless semiconductor
since the third law of thermodynamics appears to contradict the divergence of the self-energy at the chemical potential.
The ground state within the constrained Hilbert space cannot be an insulator with a complete gap nor a Mott insulator, at least when the on-site repulsion $U$ is finite. When the on-site $U$ is finite, a Mott insulator is only possible as a high temperature phase.

In the supreme single-site approximation (S$^3$A), in which the multisite self-energy is not considered, the ground state is a normal Fermi liquid. In the strong-coupling regime, the normal Fermi liquid is stabilized by the Kondo effect in the S$^3$A and is further stabilized by the Fock-type term of the superexchange interaction beyond the S$^3$A. 
The stabilization mechanism is similar to that in the resonating-valence-bond (RVB) theory for the Heisenberg model. The normal Fermi liquid stabilized by the Fock-type term is frustrated as much as an RVB spin liquid in the Heisenberg model. It is plausible that, in two dimensions and higher, the adiabatic continuation holds between an RVB spin liquid in the Heisenberg model and a normal Fermi liquid in the Hubbard model.
The Fermi liquid stabilized by the Fock-type term is a relevant {\it unperturbed\hskip2pt} state that can be used to study a normal or anomalous Fermi liquid and an ordered state in the whole Hilbert space by the Kondo lattice theory.

%%%%%%%%%%%%%%%%%%%%%%%%%%%%%%%%%%%%%%%%%%%%%%%
%%%%%%%%%%%%%%%%%%%%%%%%%%%%%%%%%%%%%%%%%%%%%%%
\appendix
\section{Proof of the Inequality (\ref{EqSelfDelta})}
\label{SecProof}
An anomalous self-energy such as one given by eq.~(\ref{EqHubbardI}), which has a pole at $\epsilon=-{\rm i}0$, can be used as a trial input for the self-energy $\Sigma_\sigma(\epsilon + {\rm i}0,{\bm k})$ at the staring point of the iterative process according to the mapping condition (\ref{EqMap-G12}) to determine the Anderson model to be solved. Thus, it is only assumed in this Appendix that $\Sigma_\sigma(\epsilon + {\rm i}0,{\bm k})$ is analytic in the upper half plane; it may have a pole or cut in the lower half plane and may be divergent on the real axis. Define the following real functions:
\begin{align}
S_1(\epsilon,{\bm k}) =
\text{Re} \left[G_{\sigma}^{-1}(\epsilon+{\rm i}0,{\bm k})\right],
\end{align}
\begin{align}
S_2(\varepsilon,{\bm k}) &= 
\text{Im} \left[G_{\sigma}^{-1}(\epsilon+{\rm i}0,{\bm k})\right],
\end{align}
\begin{align}
\tilde{S}_2(\epsilon) &= 
-\mbox{Im} \bigl[ 
\Gamma (\epsilon \!+\! {\rm i}0)
+ \tilde{\Sigma}_\sigma(\epsilon \!+\! {\rm i}0) \bigr] ,
\end{align}
\begin{align}
Y_n(\epsilon) &= \frac1{L} \sum_{\bm k}
\frac{S_1^n(\epsilon,{\bm k} )}
{S_1^2(\epsilon,{\bm k}) +S_2^2(\epsilon,{\bm k})},
\end{align}
and
\begin{align}
Z_n(\epsilon) &= \frac1{L} \sum_{\bm k}
\frac{S_2^n(\epsilon,{\bm k} )}
{S_1^2(\epsilon,{\bm k}) +S_2^2(\epsilon,{\bm k})}.
\end{align}
The site-diagonal Green function is given by
\begin{align}
R_{ii\sigma}(\epsilon+{\rm i}0) &=
Y_1(\epsilon) - i Z_1(\epsilon).
\end{align}
According to the mapping condition (\ref{EqMap-G12}), 
\begin{align}\label{EqDeltaAppendix}
\Delta(\epsilon) &=
-\text{Im} \Gamma (\epsilon \!+\! {\rm i}0) 
+\frac{\Xi(\epsilon)}
{Y_1^2(\epsilon) + Z_1^2(\epsilon)} ,
\end{align}
where
\begin{equation}
\Xi(\epsilon) = Z_1(\epsilon)-\tilde{S}_2(\epsilon)[
Y_1^2(\epsilon) + Z_1^2(\epsilon)] .
\end{equation}
In general,
\begin{equation}\label{EqPositiveS2}
S_2(\epsilon,{\bm k}) \ge 
\tilde{S}_2(\epsilon) > 0,
\end{equation}
for any ${\bm k}$.
It is trivial that
%\begin{equation}
$Y_0(\epsilon) = Z_0(\epsilon)$,
%\end{equation}
%
\begin{equation}\label{EqXYZ-1}
Y_2(\epsilon) + Z_2(\epsilon) = 1,
\end{equation}
and
\begin{equation}\label{EqXYZ-2}
Z_1(\epsilon) \ge
\tilde{S}_2(\epsilon) Y_0(\epsilon) =
\tilde{S}_2(\epsilon) Z_0(\epsilon) .
\end{equation}
According to eqs.~(\ref{EqXYZ-1}) and (\ref{EqXYZ-2}), 
\begin{align}\label{EqXiLarger0}
\Xi(\epsilon) &= 
Z_1(\epsilon)\left[Y_2(\epsilon) + Z_2(\epsilon)\right]
-\tilde{S}_2(\epsilon)[
Y_1^2(\epsilon) + Z_1^2(\epsilon)]
\nonumber \\ &\ge
\tilde{S}_2(\epsilon) \left[- Y_1^2(\epsilon) + Y_0(\epsilon)Y_2(\epsilon)\right]
\nonumber \\ & \quad 
+\tilde{S}_2(\epsilon) \left[ -Z_1^2(\epsilon) + Z_0(\epsilon)Z_2(\epsilon)
\right] .
\end{align}
Since inequalities of
\begin{equation}
\frac1{L} \sum_{\bm k}
\frac{\left[x + S_1(\epsilon,{\bm k} )\right]^2}
{S_1^2(\epsilon,{\bm k}) +S_2^2(\epsilon,{\bm k})} >0,
\end{equation}
and
\begin{equation}
\frac1{L} \sum_{\bm k}
\frac{\left[x + S_2(\epsilon,{\bm k} )\right]^2}
{S_1^2(\epsilon,{\bm k}) +S_2^2(\epsilon,{\bm k})} >0,
\end{equation}
hold for any real $x$, i.e., 
%\begin{equation}
$Y_0(\epsilon) x^2 + 2 Y_1(\epsilon) x + Y_2(\epsilon) >0$ 
%\end{equation}
and 
%\begin{equation}
$Z_0(\epsilon) x^2 + 2 Z_1(\epsilon) x + Z_2(\epsilon) >0$
%\end{equation}
%
hold for any real $x$, it follows that
\begin{equation}\label{EqY1P}
-Y_1^2(\epsilon) + Y_0(\epsilon)Y_2(\epsilon) >0,
\end{equation}
and
\begin{equation}\label{EqZ1P}
-Z_1^2(\epsilon) + Z_0(\epsilon)Z_2(\epsilon) >0.
\end{equation}
According to eqs.~(\ref{EqPositiveS2}), (\ref{EqXiLarger0}), (\ref{EqY1P}), and (\ref{EqZ1P}), it follows that $\Xi(\epsilon)>0$. Thus, the inequality (\ref{EqSelfDelta}),
$\Delta(\epsilon) \ge
-\text{Im} \Gamma (\epsilon + {\rm i}0)$,
holds as a result of eq.~(\ref{EqDeltaAppendix}) at each step of the iterative process.
 
%even if the input self-energy at the step are anomalous or divergent.

\section{Intersite Exchange Interactions in Kondo Lattices}
\label{SecKL-theory}
It is assumed in this Appendix that $U/W\gtrsim 1 $.
In general, the irreducible polarization function in spin channels, which is denoted by $\pi_s({\rm i}\omega_l,{\bm q})$, is also decomposed into the single-site $\tilde{\pi}_s({\rm i}\omega_l)$ and the multisite $\Delta\pi_s({\rm i}\omega_l,{\bm q})$:
\begin{equation}\label{EqPi-1}
\pi_s({\rm i}\omega_l,{\bm q}) =
\tilde{\pi}_s({\rm i}\omega_l) +\Delta\pi_s({\rm i}\omega_l,{\bm q}) ,
\end{equation}
where $\omega_l=2l \pi k_{\rm B}T$, with $l$ an integer, is the bosonic energy.
The single-site $\tilde{\pi}_s({\rm i}\omega_l)$ is also given by the polarization function of the self-consistently determined Anderson model.
The spin susceptibilities of the Anderson and Hubbard
models are given, respectively, by
\begin{equation}\label{EqChi-1}
\tilde{\chi}_s({\rm i}\omega_l) =
\frac{2\tilde{\pi}_s({\rm i}\omega_l) }{
1 - U \tilde{\pi}_s({\rm i}\omega_l) }, 
\end{equation}
and
\begin{equation}\label{EqChi-2}
\chi_s({\rm i}\omega_l,{\bm q}) =
\frac{2\pi_s({\rm i}\omega_l,{\bm q}) }{
1 - U \pi_s({\rm i}\omega_l,{\bm q}) }.
\end{equation}
A physical picture for a Kondo lattice is that local spin fluctuations on different sites interact by an intersite exchange interaction. According to this picture, the intersite exchange interaction $I_s({\rm i}\omega_l,{\bm q})$ is defined by
\begin{equation}\label{EqKondoKai}
\chi_s({\rm i}\omega_l,{\bm q}) =
\frac{\tilde{\chi}_s({\rm i}\omega_l) }{\displaystyle
1 - (1/4) I_s({\rm i}\omega_l,{\bm q}) \tilde{\chi}_s({\rm i}\omega_l) }.
\end{equation}
It follows from eqs.~(\ref{EqPi-1})--(\ref{EqKondoKai}) that
% (\ref{EqChi-1}), (\ref{EqChi-2}), and (\ref{EqKondoKai}) that
\begin{equation}\label{EqExch-I}
I_s({\rm i}\omega_l,{\bm q}) = 2 U^2 \Delta\pi_s({\rm i}\omega_l,{\bm q})
\Bigl\{ 1 \hskip-1pt + \hskip-1pt
O\bigl[1/U\tilde{\chi}_s({\rm i}\omega_l)\bigr]\Bigr\} .
\end{equation}
When $U/W\gtrsim 1 $, terms of $O[1/U\tilde{\chi}_s({\rm i}\omega_l)]$ can be ignored.

The exchange interaction $I_s({\rm i}\omega_l,{\bm q}) $ is composed of three terms: \cite{three-exchange,itinerant-ferro,CW1}
\begin{equation}\label{EqThreeExchange}
I_s({\rm i}\omega_l,{\bm q}) =
J_s({\bm q}) + J_Q({\rm i}\omega_l,{\bm q}) - 4\Lambda({\rm i}\omega_l,{\bm q}).
\end{equation}
The first term $J_s({\bm q})$ is the superexchange interaction, which arises from the virtual exchange of a pair excitation of electrons across the Hubbard gap.
The second term $J_Q({\rm i}\omega_l,{\bm q})$ is an exchange interaction arising from the virtual exchange of a pair excitation of  single-particle elementary excitations or quasi-particles near the chemical potential in a normal or anomalous Fermi liquid.
The third term is the remaining term, or the mode-mode coupling term among different modes of spin fluctuations, which corresponds to that in the self-consistent renormalization (SCR) theory of spin fluctuations.\cite{SCR}

When the single-site irreducible three-point vertex function in spin channels is denoted by $\tilde{\lambda}_s({\rm i}\varepsilon_l, {\rm i}\varepsilon_l+{\rm i}\omega_l; {\rm i}\omega_l)$, it follows that
\begin{align}\label{Eq3pointVertex}
\tilde{\lambda}_s(0,0;0) &= 
\tilde{\phi}_{\rm s}[1 -U \tilde{\pi}_s(0)]
\nonumber \\ &= 
\frac{2 \tilde{\phi}_{\rm s}}{U \tilde{\chi}_s(0) }
\Bigl\{ 1 +O\bigl[1/U\tilde{\chi}_s(0) \bigr] \Bigr\} ,
\end{align}
according to the Ward relation. \cite{ward} 
When $U/W\gtrsim 1 $, terms of $O[1/U\tilde{\chi}_s(0)]$ can also be ignored. 
When eq.~(\ref{Eq3pointVertex}) is approximately used, the mutual interaction mediated by intersite spin fluctuations is given by
\begin{equation}\label{EqChi-J}
\frac1{4} \bigl(U \tilde{\lambda}_s\bigr)^2
\bigl[\chi_s({\rm i}\omega_l,{\bm q}) - \tilde{\chi}_s({\rm i}\omega_l)\bigr]=
\frac1{4} \tilde{\phi}_{\rm s}^2 I_s^*({\rm i}\omega_l,{\bm q}), 
\end{equation}
where $\tilde{\lambda}_s$ represents $\tilde{\lambda}_s(0,0;0)$ and 
\begin{align}
I_s^*({\rm i}\omega_l,{\bm q}) &= 
\frac{ I_s({\rm i}\omega_l,{\bm q})}{\displaystyle
1 - (1/4) I_s({\rm i}\omega_l,{\bm q}) \tilde{\chi}_s({\rm i}\omega_l)} .
%\nonumber \\ &=
%I_s({\rm i}\omega_l,{\bm q}) + \frac1{4}I_s^2({\rm i}\omega_l,{\bm q})\chi_s({\rm i}\omega_l,{\bm q}).
\end{align}
In eq.~(\ref{EqChi-J}), the single-site term is subtracted.

%The mutual interaction mediated by intersite spin fluctuations is simply the exchange interaction $I_s^*({\rm i}\omega_l,{\bm q})$. 
%
Multisite or intersite terms can be perturbatively treated in terms of $I_s({\rm i}\omega_l,{\bm q})$ or $I_s^*({\rm i}\omega_l,{\bm q})$.
The perturbative treatment is simply the Kondo lattice theory and is also the $1/d$ expansion theory.
It should be noted that, according to eq.~(\ref{EqChi-J}), $\tilde{\phi}_{\rm s}$ is an effective three-point vertex function in this  theory. 

\section{Curie-Weiss law in a Mott Insulator}
\label{SecDefMottInsulator}
When electrons are itinerant in a magnet, the Curie-Weiss (CW) law holds for only ${\bm q}$ near a magnetic ordering wave number, which is the CW law for itinerant electron magnetism.
When electrons are localized in a magnet, the CW law holds for any ${\bm q}$, which is the CW law for local moment magnetism.
The CW law for local moment magnetism can be used to define a Mott insulator, in which electrons are localized.

First, suppose that $d\rightarrow+\infty$.
Since magnetization is a leading-order effect in $1/d$, the CW law is also a leading-order effect in $1/d$, in principle.
 In eq.~(\ref{EqThreeExchange}), $J_s({\bm q})$ and $J_Q(\omega+i0,{\bm q})$ vanish for almost all ${\bm q}$ except for particular ${\bm q}$'s such as high symmetric ${\bm q}$'s in the Brillouin zone and, if the ground state in the constrained Hilbert space is a normal Fermi liquid, nesting wave numbers in the Fermi surface; the particular ${\bm q}$'s are denoted by ${\bm Q}$. It is obvious that $- 4\Lambda(\omega+i0,{\bm q})$ vanishes for all ${\bm q}$.
Thus, the susceptibility of the Hubbard model is such that
\begin{align}
\chi_s(\omega+i0,{\bm q})=\tilde{\chi}_s(\omega+i0), 
\end{align}
for almost all ${\bm q}$ except for ${\bm Q}$ and
\begin{align}
\chi_s(\omega+i0,{\bm Q})\ne \tilde{\chi}_s(\omega+i0), 
\end{align}
for ${\bm q}={\bm Q}$.
The static susceptibility of the mapped Anderson model shows a crossover:
\begin{align}
\tilde{\chi}_s(0) = \left\{\begin{array}{cc}
\displaystyle 
\frac{c_1}{k_{\rm B} T_{\rm K}}, & T\ll T_{\rm K} \vspace{8pt}\\
\displaystyle 
\frac{c_2}{k_{\rm B} (T + c_3 T_{\rm K})}, & T\gg T_{\rm K}
\end{array}\right. ,
\end{align}
where $c_1$, $c_1$, and $c_3$ are numerical constants about unity; 
$T_{\rm K}$ may be extremely small.

% except in the case that $U/W=+\infty$ and $n=1$, in which $T_{\rm K}=0$~K.

When $k_{\rm B}T \ll U$, the $T$ dependence of $J_s({\bm Q})$ can be ignored. 
When $T\gg T_{\rm K}$, the $T$ dependence of $J_Q(\omega+i0,{\bm Q})$ can also be ignored.
When $T_{\rm K}\ll T \ll  U/k_{\rm B}$, therefore, the static susceptibility $\chi_s(0,{\bm q})$ of the Hubbard model obeys the CW law for all ${\bm q}$, the CW law for local moment magnetism.
If a Mott insulator is defined as an insulator that exhibits the CW law for local moment magnetism, it is only possible as a high temperature phase at $T\gg T_{\rm K}$ or $T\gtrsim T_{\rm K}$.

When $T\ll T_{\rm K}$ or $T\lesssim T_{\rm K}$, the $T$ dependence of $\tilde{\chi}_s(\omega+i0)$ is negligibly small; thus, that of $\chi_s(\omega+i0,{\bm q})$ is also negligibly small except for ${\bm Q}$. On the other hand, $\chi_s(\omega+i0,{\bm Q})$ can depend on $T$ because $J_Q(0,{\bm Q})$ can depend on $T$.
The precise $T$ dependence of $J_Q(0,{\bm Q})$ or $\chi_s(\omega+i0,{\bm Q})$ depends on the nature of the ground state. 
The static susceptibility $\chi_s(0,{\bm q})$ can obey the Curie-Weiss law, at least, if the ground state within the constrained Hilbert space where no order parameter exists is a normal Fermi liquid.
When the density of states for quasi-particles has a sharp peak near the chemical potential, the homogeneous one $\chi_s(0,0)$ obeys the CW law at $T>T_{\rm C}$, where $T_{\rm C}$ is the Curie temperature.\cite{CW1}
When the Fermi surface of quasi-particles shows a sharp nesting at ${\bm Q}_{\rm AF}$, the staggered one $\chi_s(0,{\bm Q}_{\rm AF})$ obeys the CW law at $T>T_{\rm N}$, where $T_{\rm N}$ is the N\'{e}el temperature.\cite{three-exchange}
The particular dependence of $\chi_s(0,{\bm q})$ on ${\bm q}$ and $T$ in infinite dimensions is a prototypic CW law for itinerant electron magnetism.
It will be interesting to study how anomalous $T$ dependence $\chi_s(\omega+i0,{\bm Q})$ can have at $T\ll T_{\rm K}$ or $T\lesssim T_{\rm K}$, if the ground state within the constrained Hilbert space is not a normal Fermi liquid.
 
In finite dimensions, according to the SCR theory of spin fluctuations, \cite{SCR}
a CW like temperature dependence is possible due to the temperature dependence of the mode-mode coupling term $- 4\Lambda(0,{\bm q})$. 
When $\rho(\epsilon)$ has no sharp peak near the chemical potential and the Fermi surface shows no sharp nesting,
$- 4\Lambda(0,{\bm q})$ has the temperature dependence that gives the CW law for almost all ${\bm q}$, but not for only ${\bm q}$ near a magnetic ordering wave number.
This dependence on $T$ and ${\bm q}$ is the so called CW law for local moment magnetism in an itinerant electron magnet. 
When $\rho(\epsilon)$ has a sharp peak near the chemical potential or the Fermi surface shows a sharp nesting, $- 4\Lambda(0,{\bm q})$ has a temperature dependence that suppresses the CW law.\cite{CW1}
This mechanism for the CW or anti-CW dependence is a higher order effect in $1/d$.

%\label{lastpage}
%%%%%%%%%%%%%%%%%%%%%%%%%%%%%%%%%%%%%%%%%%%%%%%

\begin{thebibliography}{99}
\bibitem{comMulti}
Conventional Weiss's mean-fields of magnetism or spin density wave (SDW), charge density wave (CDW), and BCS superconductivity are multisite effects but are of leading order in $1/d$.
When $U/|t|>0$, SDW can appear even in infinite dimensions.
When $U/|t|<0$, CDW or superconductivity can appear even in infinite dimensions. In the Hubbard model or a single-band model, 
other multisite effects are of higher order in $1/d$.
\bibitem{comSinglet}
For the sake of simplicity, the number of localized spins with $S=1/2$ in the Heisenberg model or electrons in the Hubbard or $t$-$J$ model 
is assumed to be even. If the number is odd, the ground state cannot be a singlet; it can be merely a doublet.
\bibitem{comTruGnd} 
The true ground state of the Hubbard model in the whole Hilbert space must be an antiferromagnetic or superconducting state in two dimensions and higher.
\bibitem{comEntropy}
When $U/|t|=+\infty$, $N<L$, 
%, the number of electrons $N$ is smaller than that of unit cells $L$ such as $N\le L$, 
and transfer integrals are nonzero only between nearest neighbors in one dimension, any eigenstate is a direct product of a charge part of $N$ spinless free fermions and a spin part of $N$ localized free spins; thus, a complete charge-spin separation occurs.
The entropy of the spin part is $N k_{\rm B}\ln 2$, even in the ground state. The static magnetic susceptibility $\chi_s(0,{\bm q})$ shows the Curie law for any ${\bm q}$ at $T\ge 0$~K although the charge part is itinerant or the ground state is a metal.
\bibitem{comAssumption}
Since $\Gamma(\epsilon+i0)=0$ corresponds to the canonical ensemble, 
the relevance of this supposition itself is doubtful in the formulation for the grand canonical ensemble.
If the supposition is irrelevant, 
what is examined here is whether or not an anomalous single-site self-energy can be a self-consistent solution as a mathematical problem.
\bibitem{comBeyond}
When single-site terms in the S$^3$A, the dynamical mean fields in DMFT, or the dynamical coherent potential in DCPA is self-consistently considered with the Fock-type term, the ground state is a singlet or a normal Fermi liquid in each approximation; no Mott metal-insulator transition occurs, even if the reservoir term is ignored.
\end{thebibliography}
\end{document}